\pdfoutput=1

\documentclass[11pt]{article}

\usepackage{ACL2023}

\usepackage{times}
\usepackage{latexsym}
\usepackage{amsmath}
\usepackage{booktabs} 
\usepackage{graphicx} 
\usepackage{array} 

\usepackage[T1]{fontenc}

\usepackage[utf8]{inputenc}

\usepackage{microtype}

\usepackage{inconsolata}

\title{Users Favor LLM-Generated Content—Until They Know It's AI}

\author{
  Petr Parshakov \\ HSE University and\\Moscow School of\\Management SKOLKOVO \\\And
  Iuliia Naidenova \\ HSE University \\\And
  Sofia Paklina \\ HSE University \\\And
  Nikita Matkin \\ HSE University \\\And
  Cornel Nesseler \\ University of Stavanger
}

\begin{document}

\maketitle

\begin{abstract}
In this paper, we investigate how individuals evaluate human and large langue models generated responses to popular questions when the source of the content is either concealed or disclosed. Through a controlled field experiment, participants were presented with a set of questions, each accompanied by a response generated by either a human or an AI. In a randomized design, half of the participants were informed of the response's origin while the other half remained unaware. Our findings indicate that, overall, participants tend to prefer AI-generated responses. However, when the AI origin is revealed, this preference diminishes significantly, suggesting that evaluative judgments are influenced by the disclosure of the response's provenance rather than solely by its quality. These results underscore a bias against AI-generated content, highlighting the societal challenge of improving the perception of AI work in contexts where quality assessments should be paramount.
\end{abstract}

\section{Introduction}

The rapid evolution of large language models (LLMs) over recent years has fundamentally transformed the landscape of text generation and human-computer interaction. These models are now integral to various applications, ranging from customer service and content creation to persuasive messaging and personalized communication. Recent advancements in LLMs have demonstrated significant potential to enhance labor productivity across various business applications, including customer communication and content creation \citep{Ayers2023-fe, Brynjolfsson2025-ne, Zhang2023-jz}. LLMs are also useful in creating persuasive public messages \citep{Karinshak2023-fg} or maintaining personalized in-depth conversations to change individual’s beliefs \citep{costello2024durably}. Understanding how individuals perceive AI-generated responses is essential for ensuring the effectiveness and acceptance of these technologies in business and social settings.

As LLM-generated content increasingly mirrors human-authored text in fluency and coherence, the challenge of distinguishing between the two sources has become more complex. Prior research has highlighted measurable differences in linguistic features and sentiment expression between human and AI texts, while also noting that the perceived quality of responses may shift when the origin of the content is disclosed. This dynamic is particularly important in settings where trust and credibility are paramount, such as in customer interactions or public communications. Many studies focus on specific areas (e.g., health or public announcements) and neglect the importance of general-interest texts.

Our study investigates how responses to popular questions on platforms such as Quora and Stack Overflow are perceived if the response comes from a human or from an LLM.  We examine these perceptions across a variety of domains, including Physical Sciences, Life Sciences, Health Sciences, Social Sciences, and Humanities, using a diverse set of popular questions. Incorporating respondent characteristics such as gender, age, educational background, and programming skills, our research aims to offer a nuanced understanding of the factors that drive trust and preference in content generation.

\section{Literature review}

\subsection{Human and LLM generated texts}
Recent research has systematically examined the distinctions between human-generated and AI-generated texts, revealing measurable differences in sentence structure, emotion expression, and other linguistic features \citep{Munoz-Ortiz2024-hp,Nitu2024-fx}. Early studies demonstrated that traditional machine learning classifiers could effectively differentiate between human and AI-generated content; however, the advent of advanced large language models (LLMs) has significantly complicated this task \citep{Hayawi2024-ri}. In fact, AI-generated texts have at times matched or even exceeded human-written texts in specific applications, such as persuasive messaging \citep{Karinshak2023-fg} and providing writing feedback in education \citep{Escalante2023-jp}. 
The increasing sophistication of LLMs has led to a significant convergence between AI-generated text and human-written content, rendering the distinction between the two increasingly challenging \citep{Ollivier2023-hx, Hayawi2024-ri}. As these models evolve, they produce text that not only mimics human writing styles but also adheres to the nuances of language, context, and coherence that characterize authentic human communication. Evaluations of existing LLM-generated text detectors have reported inconsistencies \citep{Weber-Wulff2023-lo} and high false positive rates when these systems are applied to human-authored texts \citep{Elkhatat2023-an}.
In addition, the objectivity of AI-generated content is also questionable. The literature has revealed inherent biases within outputs produced by LLMs. Studies have documented significant gender and racial biases, notably in depictions of healthcare professionals and surgeons, where male representations are frequently favored \citep{Menz2024-jp, Cevik2024-jd}. Political bias has also been observed, with certain platforms such as ChatGPT exhibiting a tendency toward left-leaning perspectives \citep{Motoki2024-iq, Rozado2024-nv}. Moreover, LLMs tend to manifest human-like content biases, as demonstrated by transmission experiments \citep{Acerbi2023-xw} and linguistic analyses \citep{Fang2024-oe}. 

\subsection{Perception of LLM-generated content}
Comparative studies reveal that human evaluators often struggle to reliably differentiate between AI-generated and human-authored content \citep{Boutadjine2024-oz}. \citet{Zhang2023-jz} demonstrated that generative and augmented AI content is frequently perceived as superior to that produced by human experts, even when humans utilize AI tools. However, disclosing the source of content narrows the perceived quality gap, suggesting a bias favoring human contributions over AI. Participants rated content more favorably when attributed to human experts, whereas awareness of AI involvement had minimal impact on perceptions. \citet{Ayers2023-fe} examined the ability of an AI chatbot (ChatGPT) to deliver quality and empathetic responses to patient questions compared to physicians. Their findings revealed that chatbot responses were preferred in the majority of evaluations, rated higher in quality, and deemed more empathetic than those of physicians. Notably, chatbot-generated texts were also significantly longer than physician responses. \citet{Karinshak2023-fg} highlighted that large language models (LLMs), particularly GPT-3, can produce high-quality persuasive content; however, individuals tend to prefer public health messages originating from human institutions rather than AI sources. 
Some studies highlight the nuanced perceptions and preferences surrounding AI-generated content across various domains. For example, \citet{Escalante2023-jp} compared human tutor feedback with AI-generated feedback in educational settings, revealing mixed results. While face-to-face interactions with tutors enhanced student engagement, AI-generated feedback was favored for its clarity and specificity. The research by \citet{Chen2024-qy} demonstrates that consumers prefer AI-generated ads with agentic appeals, while favoring human-created ads with communal appeals. 
These findings underscore that contextual factors and the awareness of text origin play a critical role in shaping user preferences.

\section{Methodology and design}

\subsection{Data collection}

This study aims to evaluate responses that are of interest to a broad audience. To achieve this, the analysis focuses on popular questions from Quora, a platform where users can post questions and provide answers, with the most popular responses prominently displayed. Additionally, for the domain of Physical Sciences and Engineering, questions from Stack Overflow, a widely used platform for programming-related queries, are incorporated. The selected questions are broadly categorized into five scientific areas: Physical Sciences and Engineering, Life Sciences, Health Sciences, Social Sciences, and Humanities. This categorization ensures balanced representation across domains and prevents the dominance of any single area. While the questions are not strictly scientific, they are designed to appeal to a general audience with diverse backgrounds. Examples of such questions include, "What started WWII?", "What was the best team in the history of sports?", and "What role does therapy play in treating anxiety?". 

From Quora and Stack Overflow, five questions are selected for each of the five scientific areas, resulting in a total of 25 questions. These questions are then posed to four prominent large language models (LLMs)—ChatGPT, Claude, Gemini, and Llama—selected for their superior performance in text generation at the time of the study. Each question generates five responses: four from the LLMs and one from a human respondent. The complete list of questions, organized by field, is presented in the Appendix~\ref{sec:appendixA1}. The average length of human responses is 1,515 characters or 265 words, while responses from the LLMs vary between 1,854 and 2,265 characters. Detailed summary statistics regarding response lengths, measured in both characters and words, are provided in the Appendix ~\ref{sec:appendixA2}.

We created a survey in which participants are randomly presented with 5 questions, each accompanied by two responses: one generated by an LLM and one by a human. Participants are also asked to provide demographic information, including age, gender, and educational background. Prior to commencing the survey, participants are informed that they will be required to choose between two responses for each question. Importantly, a random part of the participant is not informed about the origin of the responses (i.e., whether they are generated by an LLM or a human) to mitigate potential biases, such as the Hawthorne effect \citep{sedgwick2015understanding}. The experiment is registered on the Social Science Registry\footnote{\href{https://www.socialscienceregistry.org/trials/HIDDEN-FOR-REVIEW}{Social Science Registry}}. All participants were informed about how their data would be used in this study, and explicit consent was obtained prior to participation. We contacted our institution's Ethical Committee and were informed that, as we do not store any personal data, additional ethical approval was not required.

The survey is implemented in two formats: a Telegram bot and a web application developed using Streamlit. Both platforms utilize the same algorithmic structure and share an identical database of questions. In total, the study involves 993 participants, with 507 respondents using the Telegram bot and 486 the web application. From this initial pool, 130 respondents are excluded from the final sample due to their failure to provide any responses.

The survey is distributed through Prolific.com, a platform designed to match surveys with appropriate respondents. Participants are compensated at a rate of 9 GBP (approximately 11 USD) per hour. The final sample consists of 846 participants, who take an average of 6.6 minutes to complete the survey. The demographic composition of the sample is 35\% male and 45\% female, with an average age of 30 years. Additional detailed summary statistics, including breakdowns by awareness groups, are provided in the Appendix~\ref{sec:appendixB1} and the Appendix ~\ref{sec:appendixB2}.

\subsection{Estimation strategy}
\setlength{\parskip}{0pt}
To examine whether individuals prefer human- or AI-generated answers based on their awareness of the source, we employed a logistic regression analysis. Specifically, we restructured our dataset such that the unit of observation is a respondent-answer pair. We then introduced a dummy variable, \textit{human}, which takes the value of 1 if the respondent selected the human-generated answer. This variable serves as the dependent variable.\\
Among the independent variables, the primary variable of interest is \textit{aware}, which indicates whether a respondent can see the source of the answer (human or AI model). The coefficient associated with \textit{aware} reflects whether knowledge of the answer’s source influences the probability of selecting a human-generated response. Additionally, we included respondent characteristics such as gender, age, level of education, field of study, and programming skills. We also accounted for the duration of the survey and the field of the question. Furthermore, the basic model specification is presented below:
\setlength{\parskip}{0pt}
\begin{align}
    \text{Pr}(\text{human}_i = 1) = &\beta_0 + \beta_1 \text{aware}_i + \beta_2 \text{female}_i  \notag \\
    &+ \beta_3 \text{age}_i  + \beta_4 \text{duration}_i \notag \\
    &+ \beta_5 \text{education\_level}_i  \notag \\
    &+ \beta_6 \text{education\_field}_i + \notag \\ &+ \beta_7\text{programming\_skills}_i  \notag \\
    &+ \beta_8 \text{model}_i + \beta_{9} \text{question\_field}_i
\end{align}

To ensure robustness, we clustered the standard errors by gender. To assess the varying effects of \textit{aware} variable across different respondent and question characteristics, we also introduced interaction terms with gender, programming skills, and question field.

\section{Empirical results}

\subsection{Exploratory analysis}

To illustrate the possible factors that might influence the choices between answers generated by humans and AI models, we performed a graphical analysis. The most noticeable and interesting results were found in the differences in choices across gender and programming skills.

\begin{figure}
\centering
        \includegraphics[totalheight=8cm]{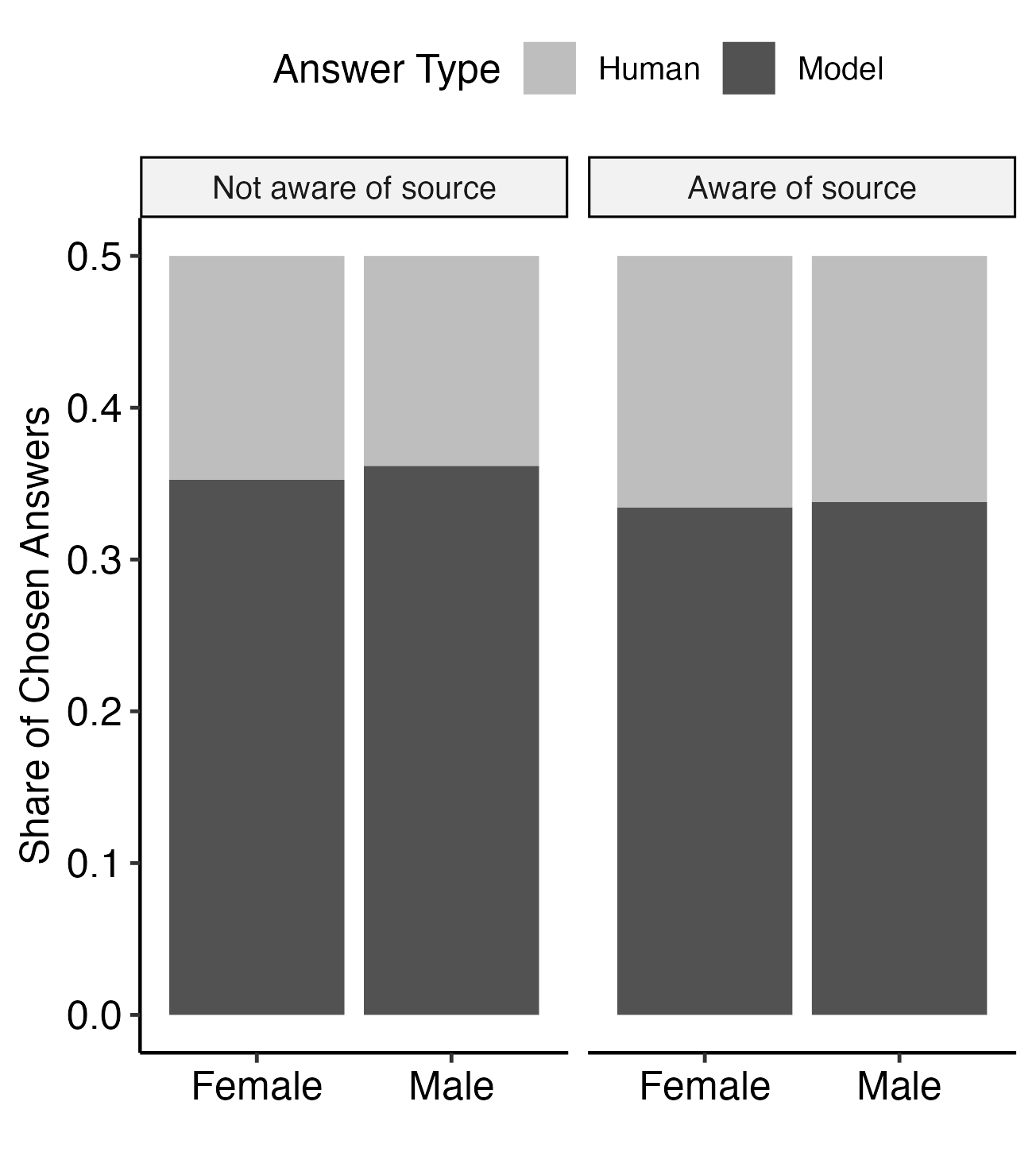}
    \caption{Distribution of chosen answers by gender and source awareness}
    \label{fig:by_gender}
\end{figure}

Figure~\ref{fig:by_gender} presents the distribution of chosen answers based on whether participants were aware of the source of the response and their gender. Across all conditions, a consistent pattern emerges, indicating a higher preference for AI-generated responses compared to human-generated ones. An interesting result is that when respondents are aware of the source of the answers, they slightly tend to choose human-generated answers, especially among male respondents.

\begin{figure}
\centering
        \includegraphics[totalheight=8cm]{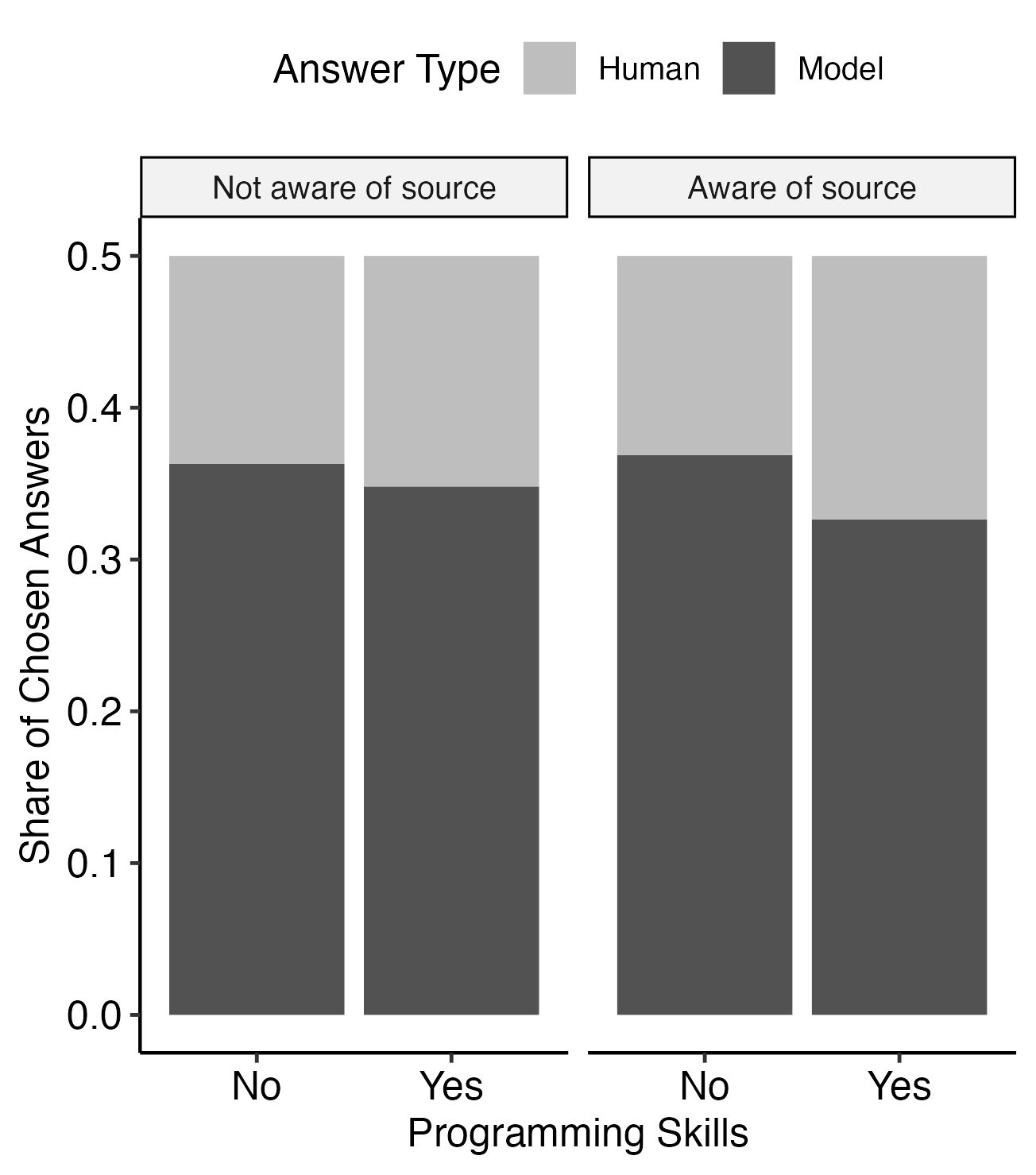}
    \caption{Distribution of chosen answers by programming skills and source awareness}
    \label{fig:by_prog_skills}
\end{figure}

Figure~\ref{fig:by_prog_skills} illustrates the distribution of selected answers based on participants’ programming skills and their awareness of the source of the responses. The results also indicate a consistent preference for AI-generated responses across all conditions. When participants were informed about the origin of the answers, the preference for AI-generated responses remained dominant, although a slight increase in the selection of human-generated responses can be observed, particularly among those with programming skills. However, this difference appears marginal and would require more in-depth hypothesis testing.

Taken together, these findings indicate that prior technical knowledge does not significantly impact the likelihood of selecting AI-generated responses. Even when participants are made aware of the source, the preference for AI-generated answers remains strong, reinforcing the notion that such responses are perceived as equally or more reliable compared to human-generated alternatives.

\subsection{Regression analysis}

Table ~\ref{tab:reg_results} presents the regression results examining individuals’ preferences for human-generated answers based on their awareness of the answer source. The dependent variable in all models is the selection of a human-generated response. Control variables for education level, field of study, programming skills, AI model, and question field are included in all models. 

\begin{table*}[!htbp] \centering  
  \fontsize{10pt}{10pt}\selectfont
\begin{tabular}{@{\extracolsep{5pt}}lcccc} 
\\[-1.8ex]\hline 
\hline \\[-1.8ex] 
 & \multicolumn{4}{c}{\textit{Dependent variable: }} \\ 
\cline{2-5} 
\\[-1.8ex] & \multicolumn{4}{c}{Ai generated answer (=0), human generated answer (=1)} \\ 
\\[-1.8ex] & (1) & (2) & (3) & (4)\\ 
\hline \\[-1.8ex] 
 Aware of AI-answer & 0.109$^{**}$ & 0.146$^{***}$ & $-$0.112 & 0.206$^{***}$ \\ 
  & (0.051) & (0.017) & (0.071) & (0.024) \\ 
  & & & & \\ 
 Female & 0.047$^{***}$ & 0.090$^{***}$ & 0.046$^{***}$ & 0.047$^{***}$ \\ 
  & (0.015) & (0.024) & (0.013) & (0.015) \\ 
  & & & & \\ 
 Age & $-$0.012$^{**}$ & $-$0.012$^{**}$ & $-$0.012$^{**}$ & $-$0.012$^{**}$ \\ 
  & (0.005) & (0.005) & (0.005) & (0.005) \\ 
  & & & & \\ 
 Duration & 0.0005 & 0.001 & 0.002 & 0.0004 \\ 
  & (0.003) & (0.003) & (0.004) & (0.003) \\ 
  & & & & \\ 
 Aware of AI-answer $\cdot$ Humanities &  &  &  & $-$0.111$^{***}$ \\ 
  &  &  &  & (0.017) \\ 
  & & & & \\ 
 Aware of AI-answer $\cdot$ Life Sciences &  &  &  & $-$0.130$^{**}$ \\ 
  &  &  &  & (0.061) \\ 
  & & & & \\ 
 Aware of AI-answer $\cdot$ Physical Sciences and Engineering &  &  &  & $-$0.068 \\ 
  &  &  &  & (0.244) \\ 
  & & & & \\ 
 Aware of AI-answer $\cdot$ Social Sciences &  &  &  & $-$0.184 \\ 
  &  &  &  & (0.125) \\ 
  & & & & \\ 
 Aware of AI-answer $\cdot$ Female &  & $-$0.089$^{***}$ &  &  \\ 
  &  & (0.023) &  &  \\ 
  & & & & \\ 
 Aware of AI-answer $\cdot$ Programming skills &  &  & 0.322$^{**}$ &  \\ 
  &  &  & (0.126) &  \\ 
  & & & & \\ 
\hline \\[-1.8ex] 
Education Level & Yes & Yes & Yes & Yes \\ 
Education Field & Yes & Yes & Yes & Yes \\ 
Programming Skills & Yes & Yes & Yes & Yes \\ 
AI Model & Yes & Yes & Yes & Yes \\ 
Question Field & Yes & Yes & Yes & Yes \\ 
Observations & 3,206 & 3,206 & 3,206 & 3,206 \\ 
Log Likelihood & $-$1,951.209 & $-$1,951.050 & $-$1,949.411 & $-$1,950.902 \\ 
\hline 
\hline \\[-1.8ex] 
\textit{Note:}  & \multicolumn{4}{r}{$^{*}$p$<$0.1; $^{**}$p$<$0.05; $^{***}$p$<$0.01} \\ 
\end{tabular} 
\caption{Logistic regression results} 
\label{tab:reg_results} 
\end{table*}  

Across all specifications, awareness of the answer source has a significant impact on individuals’ preferences. In Models (1), (2), and (4), the individual coefficient for awareness is positive and statistically significant, suggesting that when respondents know whether an answer is generated by a human or an AI, they are more likely to prefer human-generated responses.

The variable female is consistently positive and significant across all models, indicating that female respondents are more likely to prefer human-generated responses compared to male respondents. Age has a small but significant negative effect, implying that older individuals are slightly less likely to prefer human-generated answers.

The interaction effects in Models (2)–(4) provide further insights. In Model (2), the interaction between awareness of the source and female is negative and significant, indicating that the effect of knowing the source differs by gender. Since the joint effect of female and its interaction with awareness of the source is not statistically significant, we observe the positive effect of knowing the source only for male respondents. Specifically, for male respondents, knowing whether an answer is generated by a human or AI increases the probability of selecting a human-generated response.

Model (3) includes an interaction between awareness of the source and programming skills, which is positive and statistically significant, indicating that individuals with programming expertise are more likely to prefer human-generated responses when they are aware of the source.

Model (4) explores the role of the field of the question. The interaction terms show that for questions in Humanities and Life Sciences, respondents exhibit a significant negative preference for human-generated answers when they are aware of the source. However, no significant effects are observed for questions in Physical Sciences and Engineering or Social Sciences.

Overall, these findings highlight that knowing the source of an answer influences individuals’ preferences, with variations by gender, programming expertise, and the field of the question.

\section{Conclusion}

This study has investigated the influence of source awareness on individuals' preferences between human-generated and AI-generated responses. Our empirical results indicate that, although respondents generally show a tendency to favor AI-generated content, the disclosure of an answer's origin induces a measurable shift toward human-generated responses. Moreover, the extent of this shift is contingent upon respondent characteristics such as gender and programming expertise, as well as the contextual domain of the question. These findings not only extend the literature on the perceptual differences between human and AI texts \citep{Boutadjine2024-oz, Zhang2023-jz} but also complement previous studies that have noted the impact of source disclosure on content perception \citep{Ayers2023-fe, Karinshak2023-fg}.

Interestingly, women demonstrate a stronger preference for human responses compared to men, potentially due to differences in style or text length preferences, as evidenced in Appendix \ref{sec:appendixA2}. While these preferences remain consistent regardless of knowledge about the answer's origin, men exhibit a significant response to the disclosure of whether a human or an LLM authored the text. The findings on text-length differences between human and AI-generated texts align with those reported by \citet{Karinshak2023-fg}.

In contrast to earlier research that reported a pervasive difficulty among evaluators in distinguishing between AI and human-authored texts, our analysis reveals that source awareness can lead to a nuanced reallocation of trust depending on both individual and disciplinary factors. While previous studies have largely emphasized the challenges in detecting AI-generated content \citep{Weber-Wulff2023-lo, Elkhatat2023-an}, our findings suggest that transparency regarding content origin plays a critical role in shaping evaluative judgments, particularly among subgroups with specific technical proficiencies and demographic profiles.

The implications of these results are significant for the deployment of large language models in various sectors. As AI-generated content becomes increasingly prevalent in business communications, public messaging, and educational settings, ensuring that end users understand the origins of the content may enhance its perceived credibility and effectiveness. Future research should further examine the long-term effects of source transparency, extend the inquiry to encompass non-textual media, and explore the dynamics of augmented human-AI collaboration. Such investigations will be essential in refining our understanding of how best to integrate AI technologies into contexts where trust and authenticity remain paramount.

\section{Limitations}
It is important to note that our research is limited to textual content and does not encompass non-textual forms such as graphics or audio, which are widely utilized in communication. Additionally, maintaining human oversight remains crucial to ensure that generative AI-produced content is appropriate for sensitive topics and to prevent the dissemination of unsuitable material. Furthermore, our study does not explore perceptions of augmented human or augmented AI-generated content, an emerging and promising area of research \citep{Vaccaro2024-oy, Zhang2023-jz}. 

This study faces several limitations that warrant careful consideration. First, our analysis is confined to textual content and does not encompass non-textual media such as graphics, audio, or video, which are increasingly integral to modern communication. This focus on text may limit the generalizability of our findings in contexts where multimodal content plays a critical role.

Another limitation relates to our data collection methods. Although we utilized platforms such as Quora and Stack Overflow to ensure a diverse set of questions and respondent backgrounds, the online nature of these sources may introduce selection biases. The sample, while substantial, might not fully capture the broader population's cultural, demographic, or technological diversity, which could influence the observed preferences and perceptions.

Moreover, our investigation into evaluative judgments of content origin relies on self-reported responses and observable choices within an experimental framework. Despite efforts to mitigate potential biases, such as the Hawthorne effect, there remains a risk of demand effects or social desirability influencing participants' selections. This methodological constraint suggests that further studies employing alternative designs or additional qualitative measures may be needed to validate our conclusions.

Additionally, while our work contributes to understanding the distinctions between human-generated and AI-generated text, it does not explore the emerging domain of augmented human or augmented AI-generated content \citep{Vaccaro2024-oy, Zhang2023-jz}. The dynamics of collaborative content creation, where human creativity is intertwined with AI assistance, present a promising avenue for future research. Similarly, the influence of non-textual elements and the integration of multimodal communication on user perceptions remain open questions.

Finally, it is important to acknowledge that the platforms and contexts in which the data were collected may impose their own norms and biases on both human and AI-generated responses. These contextual factors could affect the style, substance, and evaluative judgments of the content. Future research should aim to replicate and extend these findings across different media and cultural settings to enhance our understanding of how source awareness influences content perception in a broader spectrum of communication environments.

\section{Ethics Statement}
This work adheres to the ACL Code of Ethics and complies with the ethical guidelines established for ACL 2023. In conducting this research, we ensured that all data collection processes were performed in accordance with ethical standards, including obtaining informed consent from all participants and anonymizing the collected data to protect privacy. We recognize that advances in large language models and the increasing prevalence of AI-generated content have significant societal implications. Accordingly, our study has been designed with a commitment to transparency and accountability. We have carefully considered potential ethical risks, including the propagation of biases and the misuse of AI-generated content, and have incorporated measures to mitigate these concerns. Our aim is to contribute to a better understanding of how source awareness influences content perception, while underscoring the importance of human oversight in the deployment of AI technologies.

\subsection{Potential Risks}
Potential risks associated with this research primarily arise from the misinterpretation and misuse of our empirical findings. In particular, the observed shift in user preferences when the origin of content is disclosed might be misconstrued to justify an inherent devaluation of AI-generated work, thereby reinforcing existing biases against AI systems. Such misinterpretations could lead to counterproductive decisions in both academic and industry settings, where effective integration of AI-generated content might be prematurely dismissed or overly prioritized based solely on its origin rather than on its quality or utility.

Another risk involves the generalization of findings from our controlled experimental setting to more complex real-world contexts. The simplified scenarios presented in our study may not fully capture the dynamic interactions and multifaceted considerations present in everyday human-AI communication. Over-reliance on these experimental results could potentially influence policy or operational practices without sufficient contextual understanding, thereby impacting decision-making processes in critical areas such as customer service, education, or public communications.

To mitigate these risks, we have taken care to contextualize our findings within a broader discussion of limitations and ethical considerations. Our study emphasizes the importance of human oversight and the need for a balanced approach in evaluating AI-generated content, ensuring that conclusions drawn from this work are interpreted with the necessary caution and within the appropriate scope.

\subsection{Information About Use Of AI Assistants}
We used ChatGPT for coding assistance and proofreading the text.

\section{Acknowledgements}
The research was supported by the Basic Research Program of the HSE University.

\bibliography{references}
\bibliographystyle{acl_natbib}

\appendix


\section{Q\&A Dataset}
\label{sec:appendixA}
\subsection{Questions Dataset}
\label{sec:appendixA1}
Table~\ref{tab:llm-bias} provides a comprehensive list of questions categorized by specific academic domains. The classification of these areas is derived from the main page of the ScienceDirect website, with the exception of the social sciences and humanities, which have been further subdivided into two distinct categories. The Physical Sciences and Engineering category encompasses questions related to physics and technological advancements. Given the prevalent use of language models in programming, two programming-related questions have been incorporated into this category. The Life Sciences category focuses on questions pertaining to natural phenomena, while the Health Sciences category addresses topics related to health and health-related technologies. The Social Sciences category comprises questions concerning economics and politics, whereas the Humanities category includes inquiries into philosophy, religion, history, and literature. These five categories are designed to represent the most widely discussed and contentious topics within their respective fields.

\subsection{Answers Dataset}
\label{sec:appendixA2}
Table~\ref{tab:word_summary} presents the descriptive statistics for word and symbol counts across various academic fields and language models. The table includes mean values and standard deviations (SD) for both symbols and words, categorized by field (Physical Sciences, Life Sciences, Health Sciences, Social Sciences, and Humanities) and model (Human, GPT-4o, Claude 3.5 Sonnet, Gemini 1.5 Pro, and Llama 3.1 405b). The data reveal variations in response length and complexity, with GPT-4o and Llama 3.1 405b generally producing longer outputs in terms of both symbols and words compared to other models. Notably, the Health Sciences field exhibits the highest symbol and word counts across most models, while the Humanities field tends to have shorter responses. These findings highlight differences in linguistic output patterns across fields and models, providing insights into the performance and characteristics of each model within specific academic domains.

\section{Survey Statistic}
\label{sec:appendixB}
\subsection{Descriptive statistics for respondents}
\label{sec:appendixB1}
Table~\ref{tab:descriptive_stats1} provides descriptive statistics for the respondent database, categorized by their awareness of the answers' source. The sample consists of 445 respondents who were not aware of the source and 401 respondents who were aware. The mean age of respondents is approximately 30 years, with similar standard deviations across both groups. Survey duration averages around 6.6 minutes, indicating consistent engagement levels. A majority of respondents (67\% and 65\% in the unaware and aware groups, respectively) report having programming skills.

Gender distribution is relatively balanced, with females comprising 46\% and 44\% of the unaware and aware groups, respectively, and males representing 34\% and 36\%. Approximately 20\% of respondents in both groups did not specify their gender. In terms of education, the majority hold a Bachelor's degree or below (72\% and 70\%), followed by Master's degrees (23\% and 25\%), and a small percentage with PhDs (5.0\% and 4.5\%).

Regarding education fields, Physical Sciences \& Engineering and Social Sciences \& Economics are the most represented, accounting for approximately 30\% and 28\% of respondents, respectively. Humanities and Health Sciences follow, while Life Sciences has the lowest representation. These statistics highlight the demographic and educational diversity of the respondent pool, ensuring a broad perspective across different fields and backgrounds.
\subsection{Descriptive statistics for questions}
\label{sec:appendixB2}
Table~\ref{tab:descriptive_stats2} presents descriptive statistics for the distribution of questions based on respondents' awareness of the answers' source. The dataset includes 2,006 questions from respondents unaware of the source and 1,773 questions from those aware of the source. Among these, human answers were chosen 29\% of the time by unaware respondents and 33\% by aware respondents, indicating a slight preference for human-generated answers when the source is known.

The distribution of AI models used in generating answers is relatively balanced across both groups. Claude, Gemini, GPT, and Llama each account for approximately 24\% to 26\% of the responses, with no significant variation between the unaware and aware groups. This suggests consistent utilization of all models regardless of respondents' awareness.

Regarding question fields, the distribution is also fairly even. Health Sciences, Humanities, Life Sciences, Physical Sciences \& Engineering, and Social Sciences each represent approximately 19\% to 22\% of the questions in both groups. This balanced representation across fields ensures a comprehensive analysis of responses across diverse academic domains. Overall, the table highlights the equitable distribution of question fields and AI models, providing a robust foundation for comparative analysis based on respondents' awareness of the answers' source.

\newpage

\begin{table*}[h]
\centering
\begin{tabular}{|p{0.3\textwidth}|p{0.7\textwidth}|} 
\hline
\textbf{Question field} & \textbf{Question} \\ \hline
Physical Sciences and Engineering & Is the earth flat? \\ \hline
Physical Sciences and Engineering & How to undo the most recent local commits in git? \\ \hline
Physical Sciences and Engineering & What does the yield keyword do in Python? \\ \hline
Physical Sciences and Engineering & Why are Nokia phones stereotyped as being indestructible? \\ \hline
Physical Sciences and Engineering & What is a black hole? How can we understand it? \\ \hline
Life Sciences & Is climate change a real and pressing concern, or is it exaggerated? \\ \hline
Life Sciences & Why do bugs fly into lights? \\ \hline
Life Sciences & Why are there no large animals (mammals or the like) with six limbs? \\ \hline
Life Sciences & Who is more intelligent, an orangutan or a chimpanzee? \\ \hline
Life Sciences & What was the first cloned animal? \\ \hline
Health Sciences & What exactly is known about the origin of Covid-19? \\ \hline
Health Sciences & Do antidepressants actually cure depression or do they just trick our brain to think differently? \\ \hline
Health Sciences & Is vaping healthier than ecigarettes? \\ \hline
Health Sciences & How good is Ozempic for weight loss? What is your review? Is it really effective? \\ \hline
Health Sciences & What role does therapy play in treating anxiety? \\ \hline
Social Sciences & What was the best team of any sport ever to compete? \\ \hline
Social Sciences & How can we spot the next financial bubble? \\ \hline
Social Sciences & What's the difference between Socialism, Marxism and Communism? \\ \hline
Social Sciences & In China, how many political parties are there? \\ \hline
Social Sciences & Would a pure Neanderthal, if one were still alive, be considered a person? \\ \hline
Humanities & What is the purpose of life \\ \hline
Humanities & Which religion is right \\ \hline
Humanities & Which language is the most difficult for people to learn? \\ \hline
Humanities & What started WWII? \\ \hline
Humanities & What is the most beautiful short poem ever written? \\ \hline
\end{tabular}
\caption{Questions by fields}
\label{tab:llm-bias}
\end{table*}

\begin{table*}[!t]
\centering
\renewcommand{\arraystretch}{1.5}
\fontsize{10pt}{10pt}\selectfont
\begin{tabular*}{\linewidth}{@{\extracolsep{\fill}}lccccc}
\toprule
\textbf{} & \textbf{Human} & \textbf{GPT-4o} & \textbf{Claude 3.5 Sonnet} & \textbf{Gemini 1.5 Pro} & \textbf{Llama 3.1 405b} \\
\midrule
\textbf{All} & & & & & \\
\hspace{1em} Symbols & 1515 (828) & 2203 (588) & 1854 (651) & 1891 (480) & 2265 (455) \\
\hspace{1em} Words & 265 (149) & 333 (83) & 283 (93) & 288 (70) & 351 (60) \\
\midrule
\textbf{Physical Sciences} & & & & & \\
\hspace{1em} Symbols & 1746 (1086) & 2133 (395) & 1680 (581) & 1927 (197) & 1969 (546) \\
\hspace{1em} Words & 302 (188) & 337 (55) & 267 (93) & 301 (28) & 320 (82) \\
\midrule
\textbf{Life Sciences} & & & & & \\
\hspace{1em} Symbols & 1508 (567) & 2026 (892) & 1515 (383) & 1693 (604) & 2222 (686) \\
\hspace{1em} Words & 268 (99) & 299 (120) & 232 (50) & 256 (93) & 342 (94) \\
\midrule
\textbf{Health Sciences} & & & & & \\
\hspace{1em} Symbols & 1702 (1144) & 2551 (174) & 2609 (182) & 2057 (648) & 2422 (343) \\
\hspace{1em} Words & 306 (216) & 381 (13) & 392 (22) & 303 (85) & 373 (28) \\
\midrule
\textbf{Social Sciences} & & & & & \\
\hspace{1em} Symbols & 1647 (595) & 2254 (576) & 2279 (468) & 2040 (487) & 2492 (231) \\
\hspace{1em} Words & 280 (113) & 344 (87) & 338 (61) & 313 (80) & 372 (31) \\
\midrule
\textbf{Humanities} & & & & & \\
\hspace{1em} Symbols & 974 (671) & 2049 (731) & 1186 (356) & 1739 (421) & 2218 (303) \\
\hspace{1em} Words & 169 (113) & 304 (101) & 185 (53) & 267 (61) & 348 (40) \\
\bottomrule
\end{tabular*}
\begin{minipage}{\linewidth}
\textsuperscript{\textit{}}Mean (SD) for continuous variables\\
\caption{Descriptive statistics for word and symbol counts by field and model}
\label{tab:word_summary}
\end{minipage}
\end{table*}

\newpage

\begin{table*}[!t]
\centering
\renewcommand{\arraystretch}{1.3} 
\fontsize{10pt}{10pt}\selectfont
\begin{tabular*}{\linewidth}{@{\extracolsep{\fill}}lcc}
\toprule
\textbf{} & \textbf{Not aware of source (N = 445)} & \textbf{Aware of source (N = 401)} \\
\midrule
\textbf{Age} & 29.90 (9.66) & 30.09 (10.33) \\
\textbf{Survey duration in minutes} & 6.62 (5.35) & 6.71 (5.36) \\
\textbf{Programming skills (1 = Yes)} & 260 (67\%) & 226 (65\%) \\
\midrule
\textbf{Gender} & & \\
\hspace{1em} Female & 206 (46\%) & 175 (44\%) \\
\hspace{1em} Male & 152 (34\%) & 145 (36\%) \\
\hspace{1em} Not stated & 87 (20\%) & 81 (20\%) \\
\midrule
\textbf{Education level} & & \\
\hspace{1em} Bachelor's Degree and Below & 316 (72\%) & 280 (70\%) \\
\hspace{1em} Master's Degree & 99 (23\%) & 100 (25\%) \\
\hspace{1em} PhD & 22 (5.0\%) & 18 (4.5\%) \\
\midrule
\textbf{Education field} & & \\
\hspace{1em} Health Sciences & 54 (12\%) & 53 (13\%) \\
\hspace{1em} Humanities & 90 (21\%) & 94 (24\%) \\
\hspace{1em} Life Sciences & 34 (7.7\%) & 21 (5.3\%) \\
\hspace{1em} Physical Sciences \& Engineering & 137 (31\%) & 121 (30\%) \\
\hspace{1em} Social Sciences \& Economics & 124 (28\%) & 109 (27\%) \\
\bottomrule
\end{tabular*}
\begin{minipage}{\linewidth}
\textsuperscript{\textit{}} Mean (SD) for continuous variables; n (\%) for categorical variables\\
\caption{Descriptive statistics for respondents by awareness of answers' source}
\label{tab:descriptive_stats1}
\end{minipage}
\end{table*}

\begin{table*}[!t]
\centering
\renewcommand{\arraystretch}{1.3} 
\fontsize{10pt}{10pt}\selectfont
\begin{tabular*}{\linewidth}{@{\extracolsep{\fill}}lcc}
\toprule
\textbf{} & \textbf{Not aware of source (N = 2,006)\textsuperscript{\textit{}}} & \textbf{Aware of source (N = 1,773)\textsuperscript{\textit{}}} \\
\midrule
\textbf{Chosen human answer} & 588 (29\%) & 583 (33\%) \\
\midrule
\textbf{AI model used} & & \\
\hspace{1em} Claude & 485 (24\%) & 422 (24\%) \\
\hspace{1em} Gemini & 529 (26\%) & 443 (25\%) \\
\hspace{1em} GPT & 472 (24\%) & 464 (26\%) \\
\hspace{1em} Llama & 520 (26\%) & 444 (25\%) \\
\midrule
\textbf{Question field} & & \\
\hspace{1em} Health Sciences & 392 (20\%) & 348 (20\%) \\
\hspace{1em} Humanities & 379 (19\%) & 331 (19\%) \\
\hspace{1em} Life Sciences & 399 (20\%) & 365 (21\%) \\
\hspace{1em} Physical Sciences \& Engineering & 438 (22\%) & 393 (22\%) \\
\hspace{1em} Social Sciences & 398 (20\%) & 336 (19\%) \\
\bottomrule
\end{tabular*}
\begin{minipage}{\linewidth}
\caption{Descriptive statistics for questions by awareness of answers' source}
\label{tab:descriptive_stats2}
\textsuperscript{\textit{}}n (\%)\\
\end{minipage}
\end{table*}

\end{document}